\documentclass[tradiabstract,letter]{aa}

\usepackage{graphicx}
\usepackage{txfonts}
\usepackage{hyperref}
\usepackage{multirow}
\usepackage{natbib}

\bibpunct{(}{)}{;}{a}{}{,} 
\begin{document}

\title{A new flaring high-energy $\gamma$-ray source}

\author{	E. Bernieri
		\inst{1,2},
		R. Campana
          	\inst{3,4},
		E. Massaro
		\inst{5},
		A. Paggi
		\inst{6},
		A. Tramacere
		\inst{7}
            }
             
\offprints{enrico.bernieri@lnf.infn.it}

\institute{INFN/LNF, via E. Fermi 40, I-00044 Frascati, Roma, Italy
\and Department of Physics, University of Roma Tre,  
via della Vasca Navale 84, I-00146 Roma, Italy
\and INAF/IAPS, via Fosso del Cavaliere 100, I-00133 Roma, Italy
\and INAF/IASF-Bologna, via Piero Gobetti 101, I-40129 Bologna, Italy
\and Department of Physics, University of Roma ``La Sapienza'', 
Piazzale A. Moro 2, I-00185 Roma, Italy
\and Harvard-Smithsonian Astrophysical Observatory, 
60 Garden Street, Cambridge, MA 02138, USA
\and ISDC, University of Geneva, Chemin d'Ecogia 16, 
Versoix, CH-1290, Switzerland
}

\date{Received ..., accepted ...}
\titlerunning{A new flaring high energy $\gamma$-ray source}
\authorrunning{E. Bernieri et al.}
	
%
\abstract{
We report the detection of a new \textbf{$\gamma$}-ray source in the \textit{Fermi}-LAT sky using
a source detection tool based on the minimal spanning tree algorithm. 
The source, not reported in previous LAT catalogues but very recently observed in the X-rays and optical bands, is characterized by an increasing $\gamma$-ray activity in 2012 June--September that reached a weekly peak flux of 
 $(3.3\pm0.6)\cdot10^{-7}$ photons~cm$^{-2}$~s$^{-1}$.
A search for a possible counterpart provides indication that it can be associated 
with the radio source NVSS~J141828+354250, whose optical SDSS colours are typical of a blazar.
}

%



\maketitle

\section{Introduction}
We report here the detection of a new high-energy $\gamma$-ray source (J1418+3542) performed in the analysis of archival 
\textit{Fermi}-Large Area Telescope (LAT) sky images at energies higher than 3 GeV, using a source detection
method based on the minimal spanning tree (MST) algorithm \citep{campana08, campana12}.

The LAT experiment \citep{atwood09} is observing the entire sky in the 0.03 to $>$300 GeV band, about once every 3~hr since 2008 August 04. 
With the increase of the sky exposure it is reasonable to expect that the number of detected stationary or moderately variable sources 
increases, thanks to the improved statistics. 
Moreover, it is also possible that a number of transient sources, that exhibit a high variability (like many blazars) 
and provide a measurable signal only for a limited time interval can unexpectedly emerge in the $\gamma$-sky. 
Among these sources only $\gamma$-ray bursts (GRBs) and sources that exceed
the threshold set by the LAT Collaboration of a daily average flux of $10^{-6}$
photons~cm$^{-2}$~s$^{-1}$ are announced as a 
bright $\gamma$-ray source\footnote{\tiny\url{http://fermi.gsfc.nasa.gov/ssc/data/policy/summary.html}}.
Other transient sources can only be observed by performing a periodic and systematic data analysis 
over time intervals of weeks or months.
The new source was detected for the first time during an MST analysis of some selected regions of the sky of low Galactic diffuse emission background, 
over suitable time intervals, typically 6--12 months. 
Subsequently, the analysis was performed in a particular region of interest surrounding the source, and
extended to the entire LAT archival dataset.
This  source started its measurable activity in 2011 February and it was not detected in the first two 
years of \emph{Fermi}-LAT observations (2008 August -- 2010 August). 
It exhibited a moderate activity in 2011 June and its flux highly increased from 2012 June to September. The source was recently observed in flaring activity in November 2012
and its discovery was communicated by an Astronomer's Telegram (ATel) on December 12 \citep{atel4643, atel4645}.
After the MST detection, we investigated the new source using the \emph{Fermi} Science 
Tools\footnote{\tiny\url{http://fermi.gsfc.gov/ssc/data/analysis/software}} to obtain count maps, light curves, and test 
statistics ($TS$) values by applying the maximum likelihood (ML) method. We also searched for possible counterparts.

\section{MST detection of the new $\gamma$-ray source}

\begin{figure*}[ht]
\centering
\includegraphics[width=0.32\textwidth]{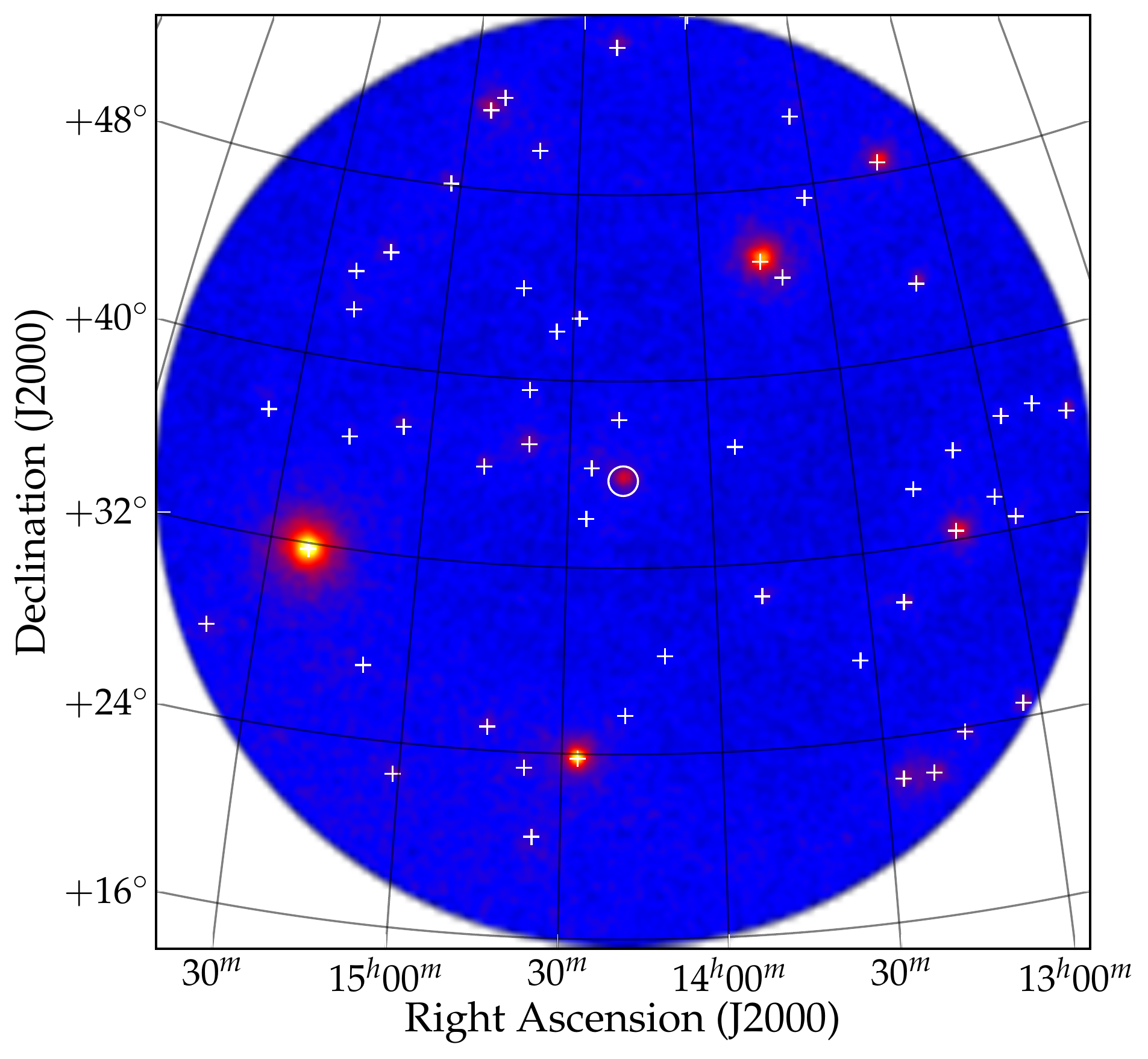}
\includegraphics[width=0.32\textwidth]{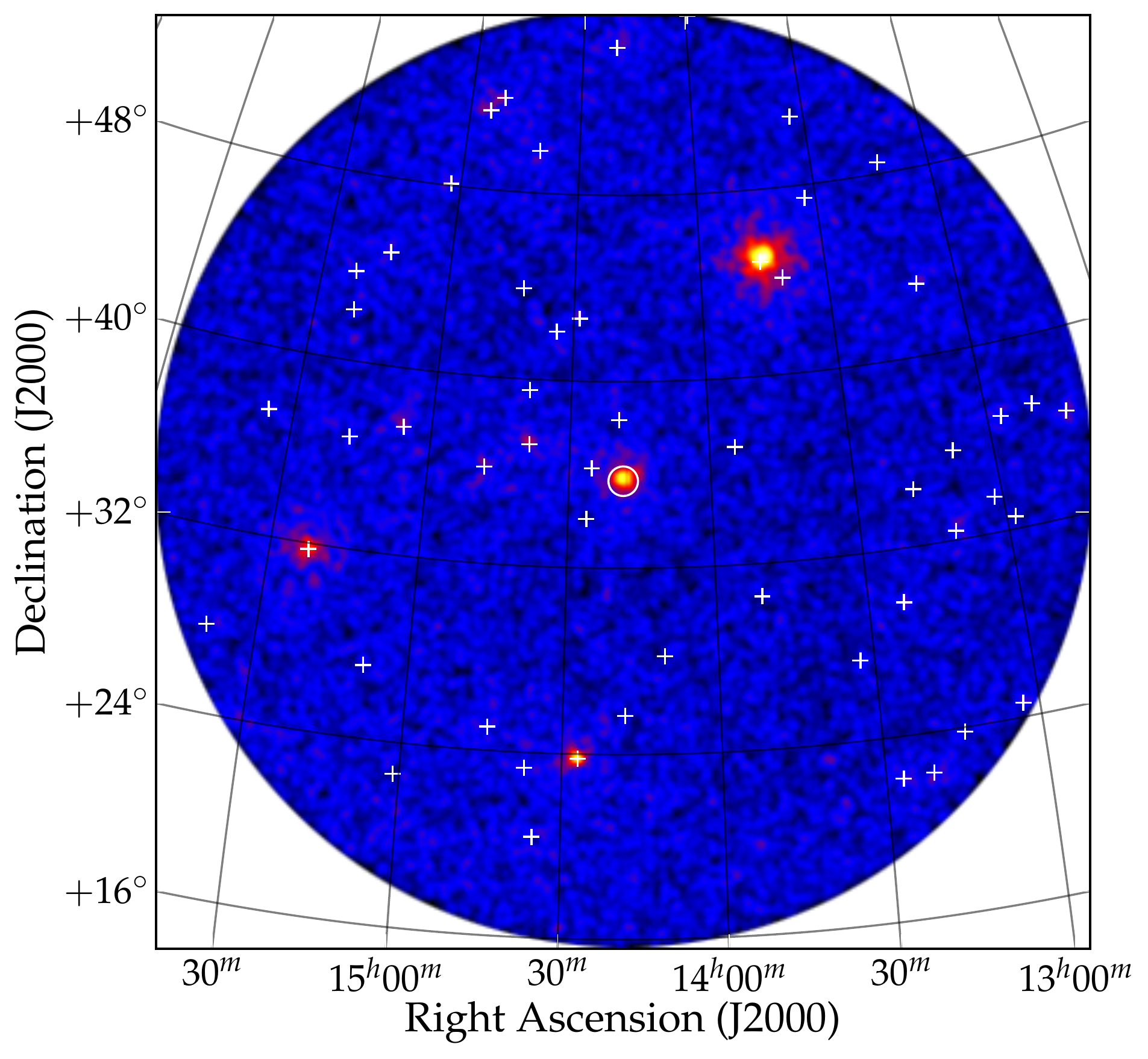}
\includegraphics[width=0.32\textwidth]{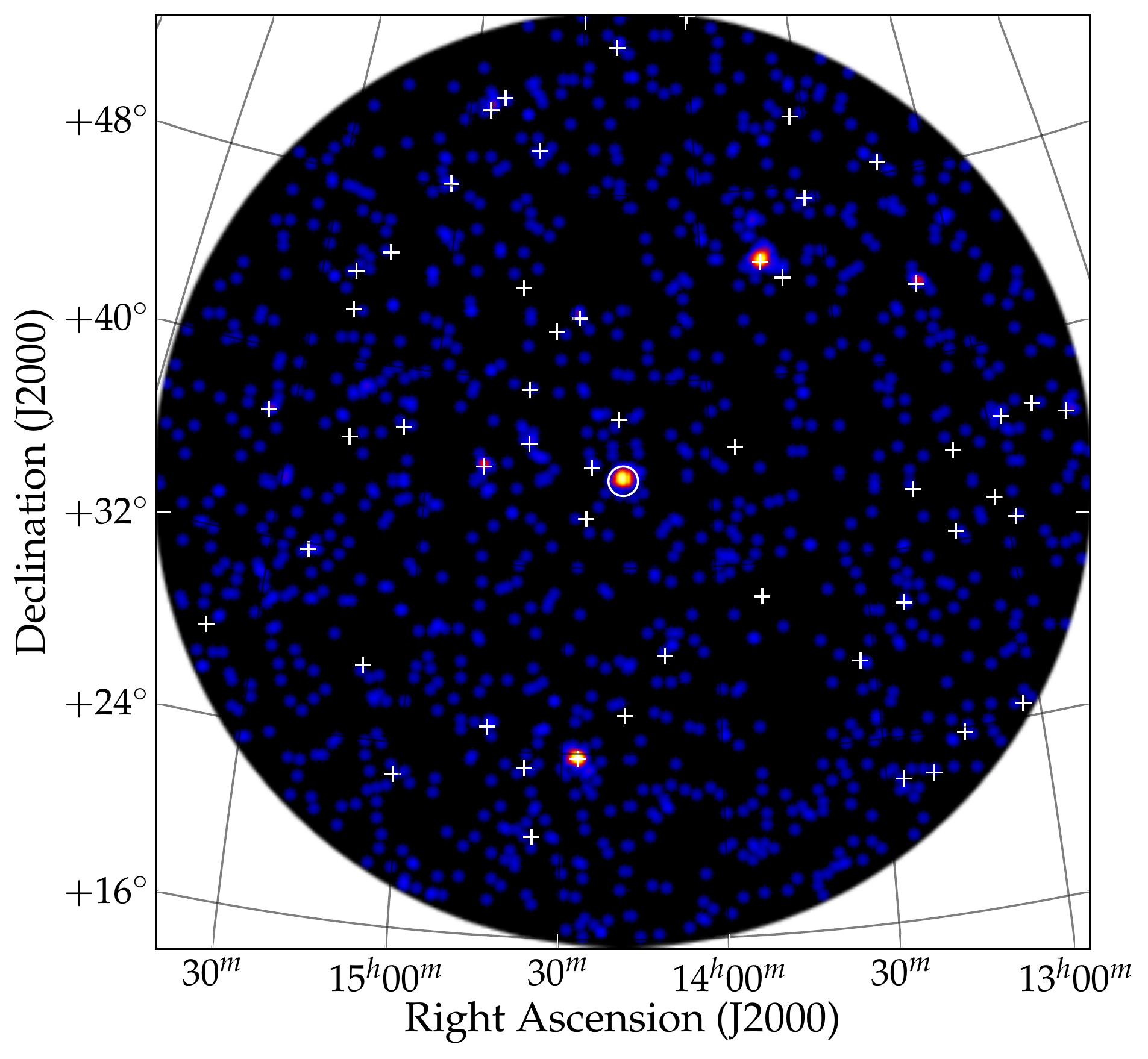}
\caption{
Photon count maps centred at the new $\gamma$-ray source  J1418+3542 
for the data from 2008 Aug 04 -- 2012 Dec 14 in the energy band 0.1--300 GeV (\emph{left panel}), 
for the data from 2012 June 01 to 2012 September 30 (flare) in the energy band  0.1--300 GeV  (\emph{central panel})
and 3--300 GeV  (\emph{right panel}).
The new source is circled at the centre of the region. White crosses mark 2FGL sources.
}
\label{figMAP}
\end{figure*}

MST is a topometric cluster-finding algorithm that exploits the pattern of ``connectedness'' 
of the detected photons, which are treated as the \emph{nodes} in a graph, where the \emph{edges} are the angular distances that
connect them.
The advantage of MST, and of other cluster-finding algorithms such as DBSCAN \citep{tramacere12}, 
is the capability to quickly find potential $\gamma$-ray sources by examining only the incoming directions of the 
photons, regardless of their energy distribution.
For our MST method we defined and successfully tested on simulated and real fields various selection 
parameters that are useful to assess the significance of the detected clusters and therefore their possible nature as genuine 
astrophysical sources \citep{campana08, campana12, massaro09}.
In particular, we found that the parameter $M$, the so-called \emph{magnitude}, which is defined as the number of photons $N$
of a cluster multiplied by its \emph{clustering degree}, i.e. the ratio of the mean edge length in the cluster to the mean value
in the field, is a very good indicator of the detection significance.
As shown by \citet{campana12}, $M$ values higher than 20 correspond to significance values higher than 4 standard 
deviations.
Detected clusters can be further analysed with other well-recognized statistical methods, such as ML 
\citep{mattox96}, to obtain an independent evaluation of their statistical significance 
and to study the time and energy properties of the source.
Our MST method was already applied to obtain lists of seed clusters for the 1FGL and 2FGL \textit{Fermi}-LAT 
catalogues \citep{abdo10a, nolan12}.

We considered all data collected by \textit{Fermi}-LAT from 2008 August 04 to 2012 December 14 in a region in the 
north Galactic emisphere, defined by a Galactic latitude $b \ge 60\degr$. 
Data were filtered using the standard LAT science tools routines \texttt{gtselect} and \texttt{gtmktime} with
a cut on the zenith angle ($<$100\degr) limb $\gamma$-rays and a cut on the rocking angle ($>$52\degr) to limit 
contamination from the Earth limb.

The MST analysis was applied to events in the 3--300 GeV energy range to avoid the low-energy background. This resulted in many clusters that were further 
filtered by applying a suitable threshold on $M$.
Most of them matched within a 0$\fdg$25 radius 2FGL sources in the region, but we also found 16 new significant clusters without any 
obvious $\gamma$-ray counterpart in LAT catalogues;
three of them have $M>35$ and one was found to have a very significant
$M = 286.8$.
The coordinates of its centre were computed by a weighted mean of the event coordinates 
with the inverse of their connecting edges \cite{campana12}.
Table \ref{table1} reports the equatorial and Galactic coordinates of the cluster centre together with the number of
events and the MST magnitude for several time intervals: active period (2011 Feb 01 -- 2012 Dec 14), pre-flare (2011 Feb 01 --  2012 May 31), 
flare (2012 Jun 01 --  2012 Sep 30), all time (2008 Aug 04 --  2012 Dec 14). 

\begin{table*}[htdp]
\caption{Main MST parameters of the source J1418+3542 in the energy range 3--300 GeV and different time intervals. The mean positional accuracy of the MST detections is 0$\fdg$06.
The maximum likelihood position was computed with the \texttt{gtfindsrc} tool with an error circle radius of 0$\fdg$02 (68\% c.l.).}
\begin{center}
\begin{tabular}{lrrrrrrl}
\hline
\hline
Time interval & RA (J2000) & Dec (J2000) & $l$ & $b$ & $N$ & $M$ & Notes \\ \hline
2008 Aug 04 -- 2011 Jan 31 &  --- &  --- &  --- &  --- & --- & --- & undetected \\
2011 Feb 01 -- 2012 Dec 14 & 214.63 &35.71 & 63.17 & 69.59 & 113 & 457.4 & active period\\
2011 Feb 01 -- 2012 May 31 & 214.59 &35.63 & 63.00 & 69.66 & 46 & 167.4 &  pre-flare\\
2012 Jun 01 -- 2012 Sep 30 & 214.67 &35.73 & 63.20 & 69.56 &  53 & 403.2 &  flare\\
2008 Aug 04 -- 2012 Dec 14 & 214.64 &35.71 & 63.16 & 69.59 &  91 & 364.4 &  full time\\ \hline
2008 Aug 04 -- 2012 Dec 14 & 214.64 &35.75 & 63.28 & 69.58 & --- & --- & ML position \\
 \hline
\end{tabular}
\end{center}
\label{table1}
\end{table*}

The source is quite evident in Figure \ref{figMAP}, left panel, which shows the event count map of a 
region of interest (ROI) of 20\degr\ radius centred at the position of the cluster. 
This ROI contains 52 2FGL sources; the new source is one of the six brightest ones in the region. 
The two other panels in Figure \ref{figMAP} show its brightening during the flare time window.

\section{Time and spectral properties}

\begin{figure}
\centering
\includegraphics[width=0.45\textwidth]{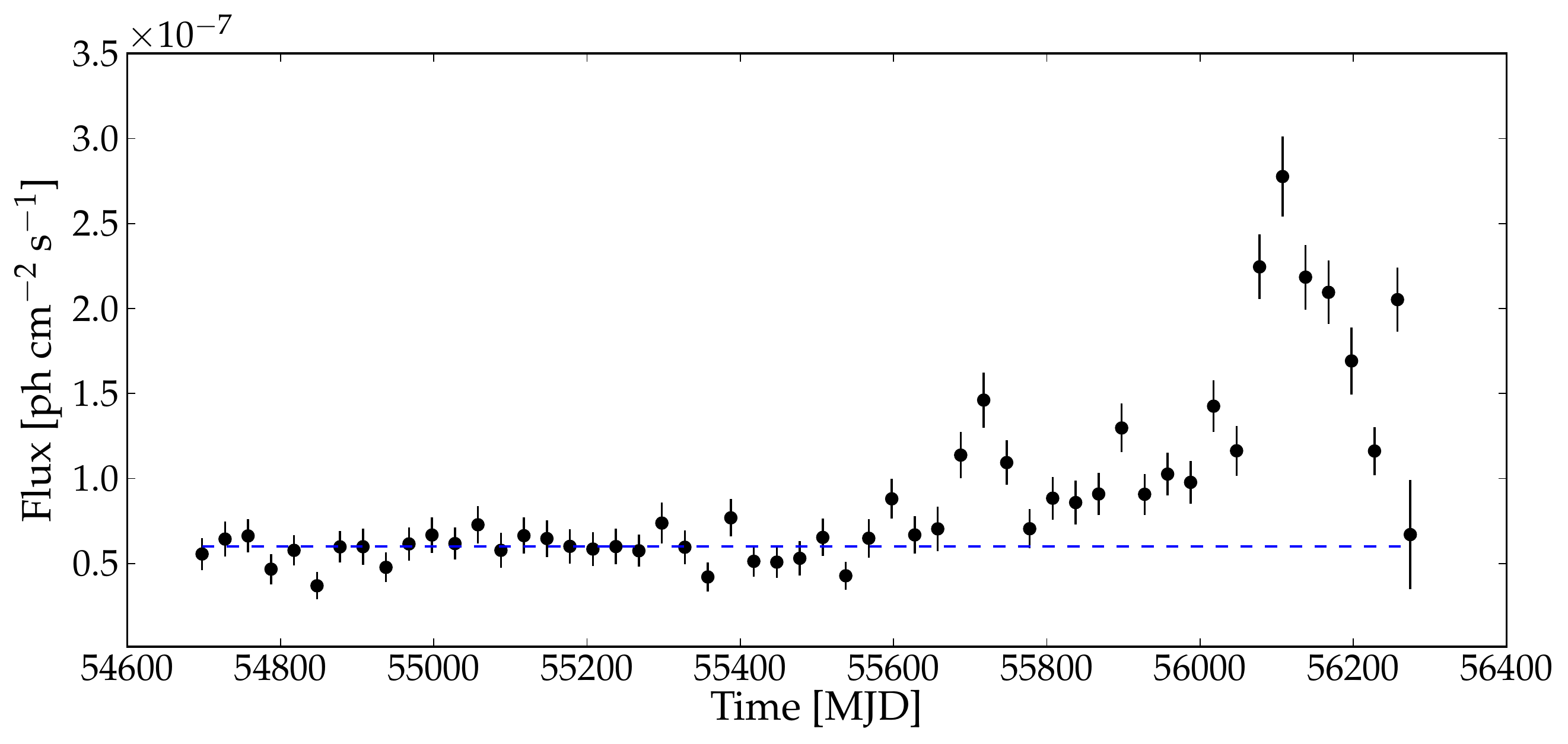}
\includegraphics[width=0.45\textwidth]{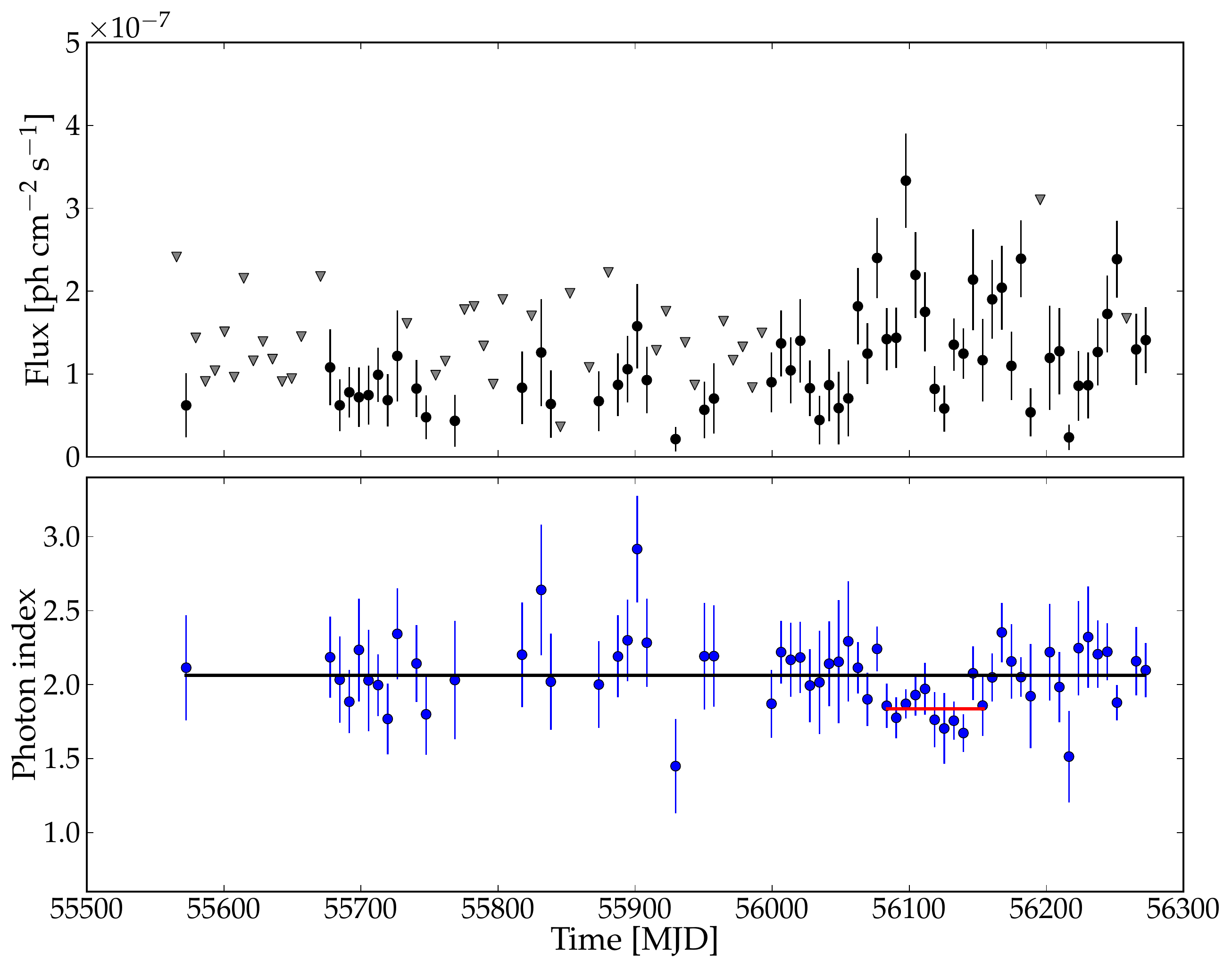}
\caption{
\emph{Top panel}: monthly aperture photometry light curve for the 0.1--100 GeV band from 2008 Aug 04 to 2012 Dec 14. 
The dashed line represents the background level.
\emph{Central panel}: weekly maximum likelihood light curve in the 0.1--100 GeV band, from 2011 Feb 01 to 2012 Dec 14. Downward triangles represent upper limits.
\emph{Bottom panel}: time evolution of the photon index; solid lines represent the mean value  over the entire
interval (black line) and during the flare (red line).
}
\label{figLC}
\end{figure}

To inspect rough features in the long-term light curve (LC) of the source we performed simple photometry 
following the standard \emph{Fermi} LAT aperture photometry 
procedure\footnote{\tiny\url{http://fermi.gsfc.gov/ssc/data/analysis/scitools/aperture_photometry.html}}. 
Figure \ref{figLC} (top panel) shows the resulting monthly LC for the entire period (2008 Aug 04 -- 2012 Dec 14) 
in the 0.1--100 GeV energy band, obtained by selecting events within a $1\degr$ aperture radius.
No model source was assumed. 
The background level was evaluated from a close spatial region of equal size without detectable sources.
Until about 2011 February (MJD$\sim$55600) the source was undetectable and remained at a quite low level until
2012 May. 
After that epoch its brightness increased and reached its maximum in about five weeks, followed by 
a decay of comparable duration.   

An ML analysis was performed on data  collected from 2011 Feb 01 to 2012 Dec 14, 
using the science tools package.
We used the \texttt{gtselect} tool to apply the cuts suggested by the 
\emph{Fermi}-LAT collaboration for point-like 
sources\footnote{\tiny\url{http://fermi.gsfc.nasa.gov/ssc/data/analysis/LAT_caveats.html}} 
to minimize the impact of the systematics and the contamination from non-photon events.  
We extracted photons with energies between 100 MeV and 100 GeV, and used the P7SOURCE\_V6 class 
event selection and the corresponding instrument response functions (IRF).

\begin{figure}
\centering
\includegraphics[width=0.45\textwidth]{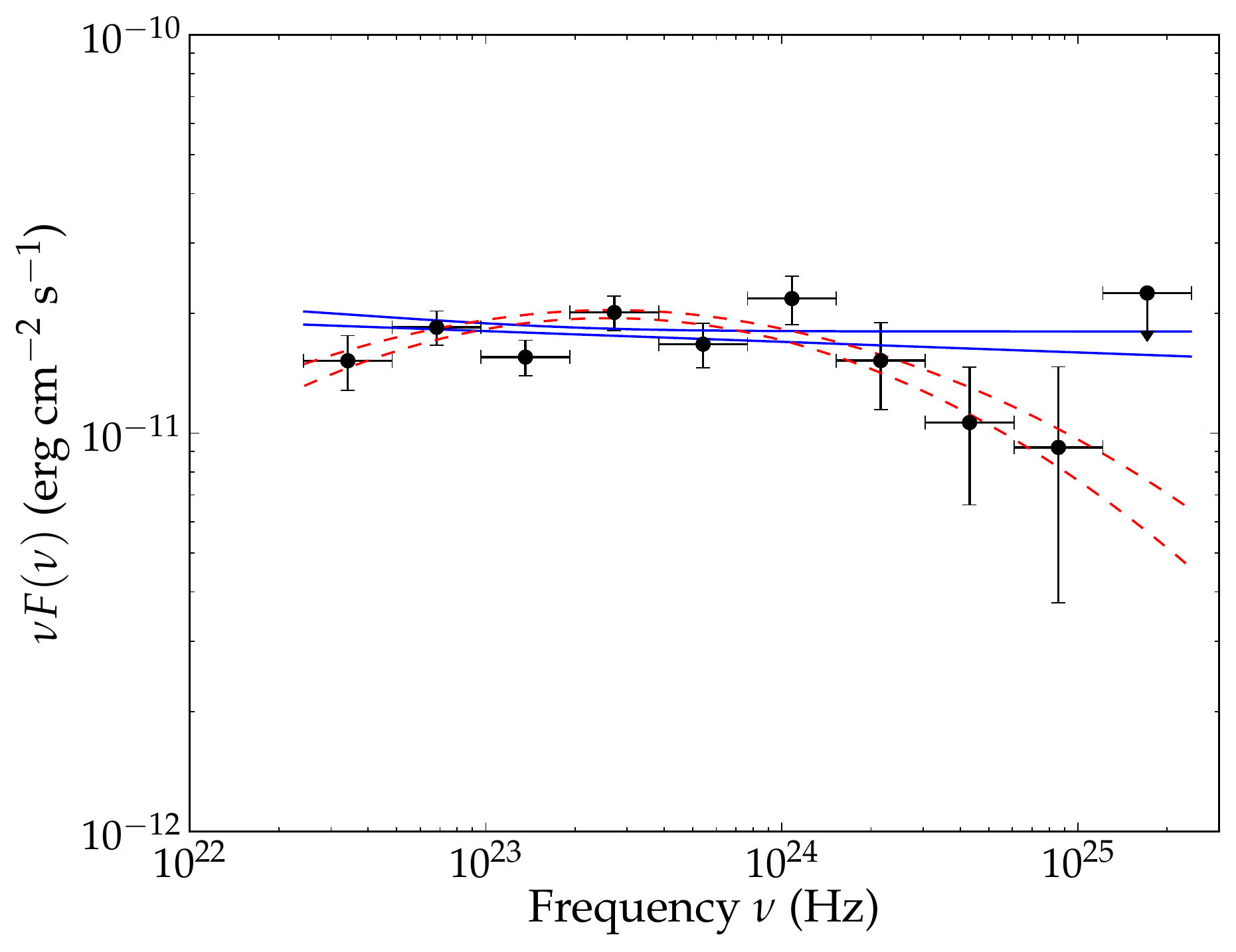}
\caption{Spectral energy distribution of the new source J1418+3542. Solid blue lines are the PL
best fit uncertainty range, while red dashed lines correspond to the LP fit.}
\label{figSED}
\end{figure}

We performed the spectral analysis using an unbinned ML estimator provided by the standard tool 
\texttt{gtlike}, and modelled each source by means of a power law (PL).
Photons were extracted from a ROI centred at the MST coordinates, within a radius 
of $10^{\circ}$. The \texttt{gtlike} region  model includes all point sources from the 2FGL catalogue
\citep{nolan12} that fall within $15^{\circ}$ from the source and a background component of the 
isotropic and Galactic diffuse emissions (standard models are available from the \textit{Fermi} Science Support 
Center\footnote{\tiny\url{http://fermi.gsfc.nasa.gov/ssc/data/access/lat/BackgroundModels.html}} (FSSC).
The ML analysis produced a weekly binned LC and a full-time spectral analysis.
Since the 2FGL catalogue provides a list of sources detected in a time window that does not overlap with 
our full-time interval, and to take into account the possible changes in the activity of the 2FGL sources in each weekly bin, 
we adopted an iterative ML procedure to obtain an optimal subtraction of the contribution
of these sources.
We selected a sub-region of the ROI with a 7$\degr$ radius: all sources of the region model  
within this radius have both the flux and the index parameter left free to vary. 
The remaining sources of the region model have the flux parameter left free and the index parameter 
frozen at the 2FGL value. A first likelihood run was performed to remove all sources with a detection significance below 3\,$\sigma$ and number of photons $<$3 from the initial region model. 
A second likelihood run was performed on the updated source list. 
Upper limits (UL) were evaluated for all sources within the 7$\degr$ radius region 
that were removed from the initial list.
The full-time spectral analysis was performed by testing a simple power law and a log-parabola (LP), 
$dN/dE \propto (E/E_0)^{-\Gamma-\beta \log(E/E_0)}$ \citep{massaro04, tramacere11}.
In the case of LP spectral law, the parameter $\beta$ measures the curvature around the peak. 
The LP distribution has only three free parameters, and the choice of the reference energy $E_0$ does 
not affect the spectral shape; we fixed its value to 300 MeV.  
We used a $TS$ based on the likelihood ratio test\footnote{The $TS$ statistics is defined as  $-$2\,Log($L_0$/$L_1$), 
where $L_0$ and $L_1$ are the maximum likelihood
estimated for the null and alternative hypothesis, respectively.}\citep{mattox96} 
to quantify the detection significance, and to check the PL model (null hypothesis) against the LP 
model (alternative hypothesis).
In the former case, the $TS$ is provided directly by \texttt{gtlike}, and the corresponding detection significance 
given by the $\sqrt{TS}$ is a $\approx$ 42\,$\sigma$ confidence level (c.l.), a very robust detection. 
The $TS$ for the comparison between the PL and the LP models returns a value of about 6, corresponding to about a 
$2.5\,\sigma$ c.l.,  meaning that we have no statistical evidence of a curved spectral shape.
The same result is obtained using a broken power law spectral model instead of an LP.
The resulting weekly LC is plotted in the central panel of Fig. \ref{figLC} and the corresponding 
photon index evolution is given in the bottom panel. 
For the cases were the analysis returned flux upper limits, we did not plot any photon index value.
The fluctuations were around the mean value of $\sim$2, with the only possible exception of the flaring period (from MJD 56080 to MJD 56150) when it was slightly harder, around $\sim$1.8. 
In any case, the photon index returned by the likelihood analysis can be biased  when the statistics is low, 
hence we cannot draw any firm conclusion about a spectral change. 
The full-time spectral energy distribution (SED) is reported in Fig. \ref{figSED}: there is a mild indication
of a steepening at high energies, although it is not statistically significant considering the rather 
large uncertainties. The best-fit parameters for the LP model are $\Gamma = 1.82 \pm 0.06$ at $E_0 = 300$ MeV, $\beta = 0.07 \pm 0.02$, 
while for the PL model they are $\Gamma = 2.03 \pm 0.03$.
The 0.1--100 GeV integral flux during the activity period is $(1.14\pm 0.06)\cdot10^{-7}$ ph~cm$^{-2}$~s$^{-1}$.
The source has a marginally significant detection in the pre-active period 2008 Aug -- 2011 Jan with a $TS = 19$\footnote{The 2FGL catalogue \citep{nolan12} used a threshold of $TS=25$ and an integration time $\sim$20\% shorter.} and a 0.1--100 GeV flux of 
$(1.1 \pm 0.4) \cdot 10^{-8}$ ph~cm$^{-2}$~s$^{-1}$. 

\section{Search for a possible counterpart}

Considering that the accuracy for the coordinates of $\gamma$-ray sources detected above a few GeV is of a few 
arcminutes, we searched for possible counterparts within a cone of radius of 10\arcmin centred at the source's
position using the ASDC sky explorer tool\footnote{\tiny\url{http://www.asdc.asi.it/}}.
There are only a few interesting objects: the radio-quiet QSO SDSS~J14186+3542 ($z=1.58$) at 2\farcm07, the X-ray 
source 1WGAJ1418.1+3543 at 4\farcm63, which are not associated with a radio and optical source, and the radio source 
GB6~J1418+3542 at the angular distance of 1\farcm23 and with a radio flux density at 4.85 GHz of about 40 mJy.
The last object appears as the most promising one. 
It is detected also at 1.4 GHz in the FIRST and NVSS catalogues which report flux densities of 52 and 61 mJy, 
respectively, and therefore its spectrum is rather flat.
Well coincident with the FIRST position is the source SDSS~J141828.58+354249.4 ($\mathrm{RA} = 214\fdg62$, $\mathrm{Dec} = 35\fdg71$) having $r=19.74$ mag and a 
colour index $u - r = 0.30 \pm 0.05$, corrected for the local interstellar reddening, which is quite low at 
this high Galactic latitude.
This blue colour index is remarkably close to that of BL Lac objects of N type \citep{mnp12} and of many flat
spectrum radio quasars (FSRQs) associated with $\gamma$-ray sources, providing a high confidence in the association 
with the new flaring $\gamma$-ray source.
This source is positionally consistent (0\farcs41 offset) with the IR counterpart listed in the Wide-field 
Infrared Survey Explorer \citep[WISE;][]{wright10} WISE~J141828.61+354249.3, which according to the WISE 
All-sky catalogue\footnote{\tiny\url{http://wise2.ipac.caltech.edu/docs/release/allsky/}}  is detected in the first 
three energy bands with the following magnitudes: 
$3.4\mbox{ $\mu$m} = 14.983 \pm 0.034$,
$4.6\mbox{ $\mu$m} = 14.215 \pm 0.042$,
$12\mbox{ $\mu$m} = 11.958 \pm 0.185$.
The three-dimensional association procedure outlined in \cite{fmassaro12a,fmassaro12b} cannot be entirely applied to 
verify if WISE J141828.61+354249.3 has IR colours consistent with those of a blazar because it is not detected at 
22~$\mu$m.
However, in the [3.4]-[4.6]-[12] \(\mu\)m colour-colour diagram, it is consistent with typical IR colours of a 
high-synchrotron-peaked BL Lac (HBL) as described in \cite{fmassaro11} for the TeV BL lac objects. 
Finally, we note that the radio source FIRST J141828.5+354249 is positionally consistent both with SDSS 
(0\farcs53 offset) and with WISE (0\farcs49 offset) sources.
Optical spectroscopic data of this source are not available and we cannot safely establish whether it is a 
BL Lac object or a FSRQ. Nevertheless its blazar nature appears to be well established and therefore it must be 
considered as the most interesting candidate for the counterpart of our new source.
The same source was also indicated by the follow-up \emph{Swift}-UVOT observations of \cite{atel4643} as the most likely counterpart.

\section{Discussion}

The high-energy $\gamma$-ray sky, according to the new scenario derived from \emph{Fermi}-LAT observations, 
appears to be dominated by thousands of sources, many of which are to be observed variable over different time scales.
Searching for transient sources is useful to investigate their behaviour and, particularly, to evaluate the typical
duration of their activity periods and the possible relations with their luminosity and spectral properties, which in turn useful for estimating their contribution to the diffuse background in
different energy bands.
The new $\gamma$-ray source, found by us by means of the MST algorithm at energies higher than 3 GeV 
with $M = 286.6$, is clearly bright enough to be detected by any method.
Its discovery was reported to the astronomical community after a new flare occurred on November 20 \citep{atel4643, atel4645}
when our analysis was already complete.
Selecting shorter time windows during the flaring period would produce a large increase
of $M$ even though the event number in the cluster is reduced, which confirms that $M$ is a good indicator of the signal-to-noise ratio. 
Table \ref{table1} shows that the positional accuracy of our MST implementation is very good, which confirms that this
clustering method is very promising for the detection and location of transient sources.

The time and spectral behaviour of J1418+3542 in the $\gamma$-ray band appear to be those of a blazar. Therefore we are confident about the
proposed counterpart; we cannot safely establish, however, whether it is a BL Lac object or a FSRQ because its
properties appears to be borderline between these two types.
Radio and $\gamma$-ray spectra are suggestive of a quasar, while the IR WISE colours are those of a HBL source.
Spectral measurements in the optical band would be very useful to establish its nature.

Finally, we underline that the brighter flare has a typical FWHM duration (5 months) of about 10\% of the 
entire observation period, suggesting that an elusive population of extragalactic sources characterized by even
shorter activity time intervals can actually exist and contribute to the isotropic background.
Efficient tools for detecting these sources are therefore very useful for a complete description of the high-energy cosmic landscape.

\begin{acknowledgements}
Part of this work is based on archival data, software or on-line services provided by the ASI Science Data Center (ASDC). 
We furthermore acknowledge use of archival Fermi and SDSS data. Funding for the SDSS and SDSS-II has been provided by the Alfred P. Sloan Foundation, the Participating Institutions, the National Science Foundation, the U.S. Department of Energy, the National Aeronautics and Space Administration, the Japanese Monbukagakusho, the Max Planck Society, and the Higher Education Funding Council for England. The SDSS Web Site is \url{http://www.sdss.org}. We are very grateful to Gino Tosti for useful discussions and to the referee for her/his insightful comments.
\end{acknowledgements}

\bibliography{bibliography} 

\begin{thebibliography}{17}
\expandafter\ifx\csname natexlab\endcsname\relax\def\natexlab#1{#1}\fi

\bibitem[{{Abdo} {et~al.}(2010){Abdo}, {Ackermann}, {Ajello}, {Allafort},
  {Antolini}, {Atwood}, {Axelsson}, {Baldini}, {Ballet}, {Barbiellini}, \&
  et~al.}]{abdo10a}
{Abdo}, A.~A., {Ackermann}, M., {Ajello}, M., {et~al.} 2010, \apjs, 188, 405

\bibitem[{{Atwood} {et~al.}(2009){Atwood}, {Abdo}, {Ackermann}, {Althouse},
  {Anderson}, {Axelsson}, {Baldini}, {Ballet}, {Band}, {Barbiellini}, \&
  et~al.}]{atwood09}
{Atwood}, W.~B., {Abdo}, A.~A., {Ackermann}, M., {et~al.} 2009, \apj, 697, 1071

\bibitem[{{Campana} {et~al.}(2012){Campana}, {Massaro}, {Bernieri}, {Tinebra},
  \& {Tosti}}]{campana12}
{Campana}, R., {Massaro}, E., {Bernieri}, E., {Tinebra}, F., \& {Tosti}, G.
  2012, in preparation

\bibitem[{{Campana} {et~al.}(2008){Campana}, {Massaro}, {Gasparrini}, {Cutini},
  \& {Tramacere}}]{campana08}
{Campana}, R., {Massaro}, E., {Gasparrini}, D., {Cutini}, S., \& {Tramacere},
  A. 2008, \mnras, 383, 1166

\bibitem[{{Dutka} {et~al.}(2012){Dutka}, {Ojha}, {Tanaka}, \&
  Sokolovsky}]{atel4643}
{Dutka}, M., {Ojha}, R., {Tanaka}, Y., \& Sokolovsky, K. 2012, The Astronomer's
  Telegram, 4643

\bibitem[{{Mahabal} {et~al.}(2012){Mahabal}, {Drake}, {Djorgovski}, {Graham},
  {Williams}, {Donalek}, {Greaves}, {Prieto}, {Catelan}, {Christensen},
  {Beshore}, \& {Larson}}]{atel4645}
{Mahabal}, A.~A., {Drake}, A.~J., {Djorgovski}, S.~G., {et~al.} 2012, The
  Astronomer's Telegram, 4645

\bibitem[{{Massaro} {et~al.}(2012{\natexlab{a}}){Massaro}, {Nesci}, \&
  {Piranomonte}}]{mnp12}
{Massaro}, E., {Nesci}, R., \& {Piranomonte}, S. 2012{\natexlab{a}}, \mnras,
  422, 2322

\bibitem[{Massaro {et~al.}(2004)Massaro, Perri, Giommi, \& Nesci}]{massaro04}
Massaro, E., Perri, M., Giommi, P., \& Nesci, R. 2004, \aap, 413, 489

\bibitem[{{Massaro} {et~al.}(2009){Massaro}, {Tinebra}, {Campana}, \&
  {Tosti}}]{massaro09}
{Massaro}, E., {Tinebra}, F., {Campana}, R., \& {Tosti}, G. 2009, in 2009 Fermi
  Symposium eConf Proceedings, Vol. C091122

\bibitem[{{Massaro} {et~al.}(2011){Massaro}, {D'Abrusco}, {Ajello}, {Grindlay},
  \& {Smith}}]{fmassaro11}
{Massaro}, F., {D'Abrusco}, R., {Ajello}, M., {Grindlay}, J.~E., \& {Smith},
  H.~A. 2011, \apjl, 740, L48

\bibitem[{{Massaro} {et~al.}(2012{\natexlab{b}}){Massaro}, {D'Abrusco},
  {Tosti}, {Ajello}, {Gasparrini}, {Grindlay}, \& {Smith}}]{fmassaro12b}
{Massaro}, F., {D'Abrusco}, R., {Tosti}, G., {et~al.} 2012{\natexlab{b}}, \apj,
  750, 138

\bibitem[{{Massaro} {et~al.}(2012{\natexlab{c}}){Massaro}, {D'Abrusco},
  {Tosti}, {Ajello}, {Paggi}, \& {Gasparrini}}]{fmassaro12a}
{Massaro}, F., {D'Abrusco}, R., {Tosti}, G., {et~al.} 2012{\natexlab{c}}, \apj,
  752, 61

\bibitem[{{Mattox} {et~al.}(1996){Mattox}, {Bertsch}, {Chiang}, {Dingus},
  {Digel}, {Esposito}, {Fierro}, {Hartman}, {Hunter}, {Kanbach}, {Kniffen},
  {Lin}, {Macomb}, {Mayer-Hasselwander}, {Michelson}, {von Montigny},
  {Mukherjee}, {Nolan}, {Ramanamurthy}, {Schneid}, {Sreekumar}, {Thompson}, \&
  {Willis}}]{mattox96}
{Mattox}, J.~R., {Bertsch}, D.~L., {Chiang}, J., {et~al.} 1996, \apj, 461, 396

\bibitem[{{Nolan} {et~al.}(2012){Nolan}, {Abdo}, {Ackermann}, {Ajello},
  {Allafort}, {Antolini}, {Atwood}, {Axelsson}, {Baldini}, {Ballet}, \&
  et~al.}]{nolan12}
{Nolan}, P.~L., {Abdo}, A.~A., {Ackermann}, M., {et~al.} 2012, \apjs, 199, 31

\bibitem[{Tramacere {et~al.}(2011)Tramacere, Massaro, \& Taylor}]{tramacere11}
Tramacere, A., Massaro, E., \& Taylor, A.~M. 2011, \apj, 739, 66

\bibitem[{{Tramacere} \& {Vecchio}(2012)}]{tramacere12}
{Tramacere}, A. \& {Vecchio}, C. 2012, \aap\ in press

\bibitem[{{Wright} {et~al.}(2010){Wright}, {Eisenhardt}, {Mainzer}, {Ressler},
  {Cutri}, {Jarrett}, {Kirkpatrick}, {Padgett}, {McMillan}, {Skrutskie},
  {Stanford}, {Cohen}, {Walker}, {Mather}, {Leisawitz}, {Gautier}, {McLean},
  {Benford}, {Lonsdale}, {Blain}, {Mendez}, {Irace}, {Duval}, {Liu}, {Royer},
  {Heinrichsen}, {Howard}, {Shannon}, {Kendall}, {Walsh}, {Larsen}, {Cardon},
  {Schick}, {Schwalm}, {Abid}, {Fabinsky}, {Naes}, \& {Tsai}}]{wright10}
{Wright}, E.~L., {Eisenhardt}, P.~R.~M., {Mainzer}, A.~K., {et~al.} 2010, \aj,
  140, 1868

\end{thebibliography}
\end{document}